\documentclass[a4paper,english,conference]{IEEEtran}
\usepackage{latexsym}
\usepackage{float}
\usepackage{amsfonts}
\usepackage{amsbsy}
\usepackage{amssymb}
\usepackage{times}
\usepackage{graphicx}
\usepackage{setspace}
\usepackage{enumerate}
\usepackage[usenames]{color}
\usepackage[dvips]{pstcol}
\usepackage{epstopdf}
\usepackage{cite}
\usepackage{amssymb}
\usepackage{amsfonts}
\usepackage{graphicx}
\usepackage{epsfig}
\usepackage{psfrag}
\usepackage{xcolor}
\usepackage{amsfonts, bm}
\usepackage{epstopdf}
\usepackage{cite}
\usepackage{color}
\usepackage{xcolor}
\usepackage{subfig}
\usepackage{verbatim}
\usepackage{multirow}
\usepackage{array}
\usepackage{booktabs}
\usepackage{amsthm}
\usepackage{makecell}
\usepackage{units}
\usepackage[linesnumbered, ruled]{algorithm2e}
\usepackage{algpseudocode}
\usepackage[ruled]{algorithm2e}
\usepackage{amsmath}

\IEEEoverridecommandlockouts
\begin{document}
\title{FAST: Feature Arrangement for Semantic Transmission}
%%%%%%%%%%%%%%%%%%%%%%%%%%%%%%%%%%%%%%%%
\author{\IEEEauthorblockN{ Kequan Zhou, Guangyi Zhang, Yunlong Cai, Qiyu Hu, and Guanding Yu}
	\IEEEauthorblockA{College of Information Science and Electronic Engineering, Zhejiang University, Hangzhou, China \\ E-mail: \{kqzhou, zhangguangyi, ylcai, qiyhu, yuguanding\}@zju.edu.cn} }

\maketitle
\vspace{-3.3em}
\begin{abstract}
Although existing semantic communication systems have achieved great success, they have not considered that the channel is time-varying wherein deep fading occurs occasionally.
Moreover, the importance of each semantic feature differs from each other.
Consequently, the important features may be affected by channel fading and corrupted, resulting in performance degradation.
Therefore, higher performance can be achieved by avoiding the transmission of important features when the channel state is poor.
In this paper, we propose a scheme of Feature Arrangement for Semantic Transmission (FAST).
In particular, we aim to schedule the transmission order of features and transmit important features when the channel state is good.
To this end, we first propose a novel metric termed feature priority, which takes into consideration both feature importance and feature robustness.
Then, we perform channel prediction at the transmitter side to obtain the future channel state information (CSI).
Furthermore, the feature arrangement module is developed based on the proposed feature priority and the predicted CSI by transmitting the prior features under better CSI.
Simulation results show that the proposed scheme significantly improves the performance of image transmission compared to existing semantic communication systems without feature arrangement.
\end{abstract}
%\vspace{-1.3em}

\IEEEpeerreviewmaketitle

\section{Introduction}
In recent years, semantic communications have emerged as novel communication schemes, which aim to transmit the semantic information behind the source data, thereby improving communication efficiency.
With the development of deep learning (DL) techniques, many DL-enabled semantic communication systems have been proposed recently \cite{DJSCC, DJSCCrate, VarLength, MemSC, SNRadapt, ADJSCC, ADJSCC-I, UDeepSC}.
Different from conventional communications, semantic communications endeavor to design source coding and channel coding jointly.
Moreover, the source data is encoded into semantic features, which contain the semantic information behind the data.
In particular, the authors in \cite{DJSCC} firstly proposed a DL-based joint source-channel coding (DJSCC) scheme for image transmission.
In light of \cite{DJSCC}, the authors in \cite{DJSCCrate} proposed an adaptive-rate scheme by selectively transmitting semantic features.
A policy network has been trained to select the features and only the important ones are transmitted.
In addition, a variable-length image compression scheme has been developed in \cite{VarLength}, which also takes feature importance into account.
Furthermore, the authors in \cite{MemSC} succeeded in reducing the overhead by masking the unimportant elements, which are recognized through training the model with mutual information.

Despite the insight into feature importance, the aforementioned works have not well investigated the impacts of physical channels.
To deal with channel impairment, channel state information (CSI) has been taken  into consideration in existing works.
Specifically, the authors in \cite{SNRadapt} developed an SNR-adaptive system for multi-user scenarios. They estimate the SNR at the receiver side and exploit it to adaptively decode the received features, which makes the system adaptable to different SNRs.
In \cite{ADJSCC}, a novel SNR-adaptive scheme has been designed by resorting to attention mechanisms.
This system is jointly trained with CSI and can operate at different SNR levels.
By following the works in \cite{ADJSCC}, the authors in \cite{ADJSCC-I} further proposed an attention mechanism-based multi-layer JSCC architecture for progressive image transmission.
Moreover, the dynamic scheme proposed in \cite{UDeepSC} can also adapt to different channel conditions and adjust the number of transmitted features accordingly.

\begin{figure}[H]
	\begin{centering}
		\includegraphics[width=0.49\textwidth]{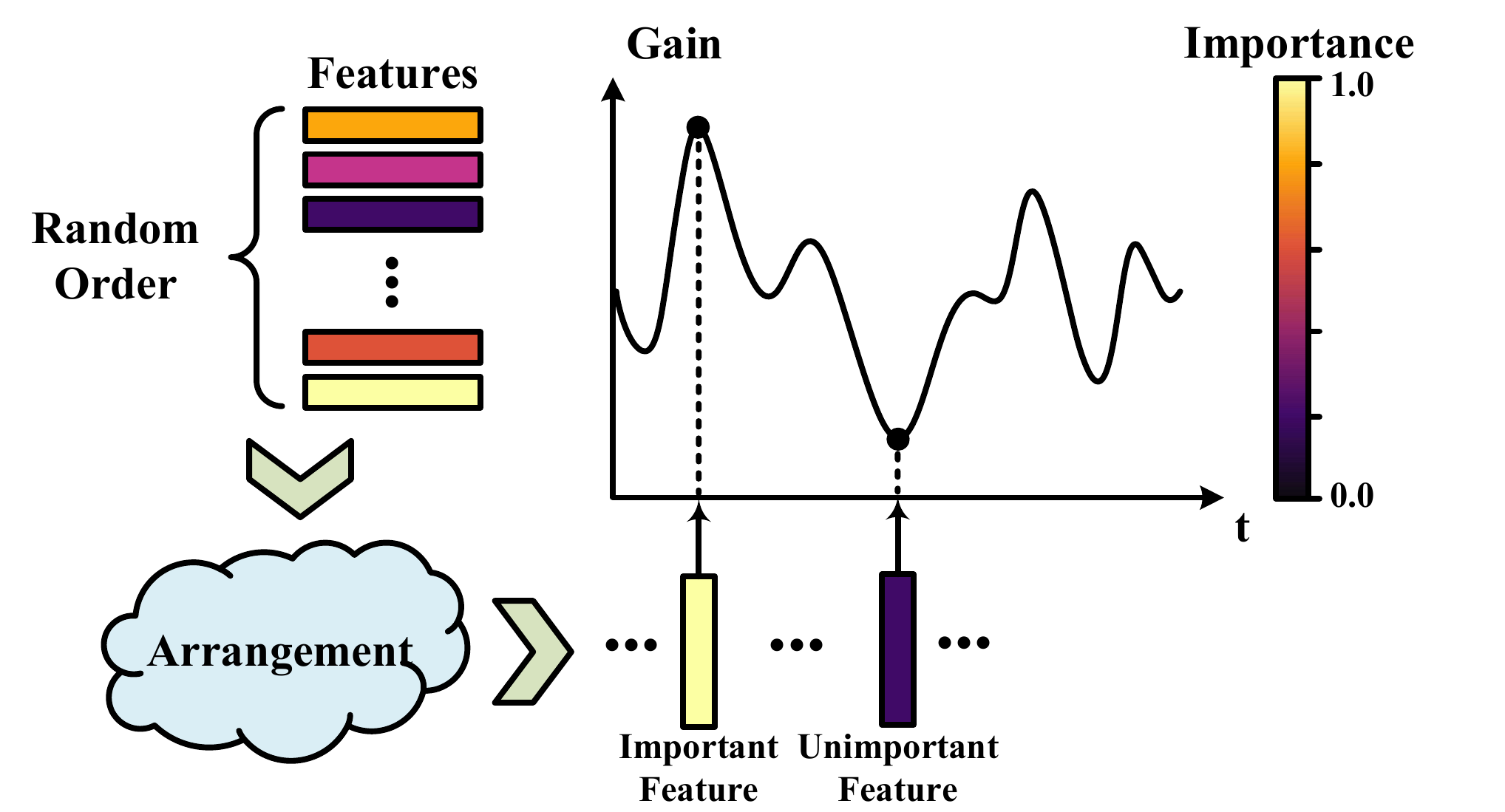}
		\par\end{centering}
	\caption{The concept of feature arrangement.}
	\label{Arrangement}
\end{figure}

\begin{figure*}[t]
	\begin{centering}
		\includegraphics[width=0.9\textwidth]{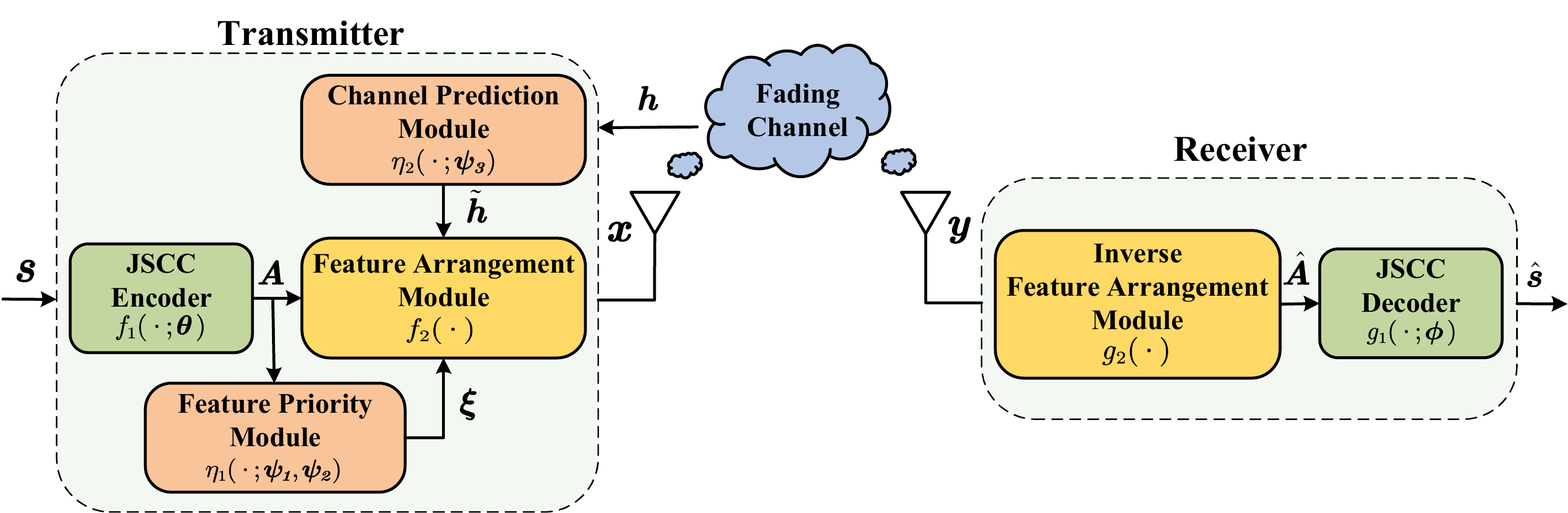}
		\par\end{centering}
	\caption{The framework of the proposed FAST.}
	\label{Framework}
\end{figure*}

Although the aforementioned works with SNR-adaptive mechanisms have exhibited excellent performance in different tasks, they only employ simple channel models that cannot characterize the time-varying feature of realistic fading channels.
Besides, the importance of each semantic feature is different from one another.
Consequently, important features may unexpectedly experience channel fading and be corrupted as CSI varies.
Therefore, as shown in Fig. \ref{Arrangement}, it is possible to achieve better performance by arranging the transmission order of the features.
Moreover, the robustness of semantic features has been studied in \cite{BoostRobust}.
Considering the transmission in fading channels, features with low robustness are easily corrupted.
This indicates that the features with high importance and low robustness have higher priority to be transmitted under good CSI than those features with high importance and high robustness.
Thus, it is worth developing a new metric to quantify feature priority, which considers feature importance and feature robustness simultaneously.

In this paper, we propose a scheme of Feature Arrangement for Semantic Transmission (FAST), whose  acronym FAST implies the efficient transmission of semantic communications, and our main contributions are summarized as follows.

\begin{itemize}
\item We propose a novel algorithm to calculate feature priority by taking into account both feature importance and feature robustness.
\item We employ the knowledge distillation technique \cite{KD} to simulate the feature priority algorithm via neural networks during transmission, which is more practical than directly performing the algorithms and significantly reduces the latency and computational overheads.
\item Based on the feature priority and predicted CSI, a feature arrangement module is designed to schedule the transmission order of the features.
This design ensures that the features with high priority are transmitted when the CSI is better, and those with low priority are transmitted when the CSI is worse, which enhances the reliability of semantic transmission.
\item Simulation results show that the proposed FAST brings remarkable performance gain compared to existing semantic communication systems without feature arrangement.
\end{itemize}

The rest of the paper is organized as follows.
Section \ref{System} introduces the framework of FAST.
Then, the details of the feature arrangement are presented in Section \ref{Methods}.
In Section \ref{Simulation}, simulation results are provided.
Finally, the paper is concluded in Section \ref{Conclusion}.

\section{Framework of FAST} \label{System}
In this section, we propose the framework of FAST.
As shown in Fig. \ref{Framework}, the FAST is composed of the JSCC encoder and decoder, the feature priority module, the channel prediction module, and the (inverse) feature arrangement module.

\subsection{Overview of FAST}
In particular, an input image is represented by a vector, $\bm{s}\in \mathbb{R}^l$, where $l$ is the size of the image.
Then, the transmitter firstly encodes $\bm{s}$ into a feature tensor, $\bm{A}\in \mathbb{R}^{c\times h\times w}$, where $c$ is the number of features and $(h\times w)$ is the shape of each feature.
The process is represented as
\begin{equation}
	\bm{A}=f_1(\bm{s};\bm{\theta}),
\end{equation}
where $\bm{\theta}$ denotes the parameter set of the encoder, $f_1(\cdot)$.

Subsequently, with the assistance of the channel prediction module and the feature priority module, the feature arrangement module arranges the order of the semantic features in $\bm{A}$.
Next, the arranged feature tensor is mapped into the channel input symbol vector, $\bm{x}\in \mathbb{C}^k$, where $k$ is the number of symbols.
Moreover, $\bm{x}$ is subject to the average power constraint, $P$, at the transmitter, i.e.,
$\frac{1}{k}||\bm{x}||^2\leq P$.

Then, the symbol vector received at the receiver is given by
\begin{equation}
	\bm{y}=h\bm{x} + \bm{n},
\end{equation}
where $h\in \mathbb{C}$ is the channel realization and $\bm{n}\in \mathbb{C}^k$ is the additive white Gaussian noise with the distribution, $\mathcal{CN}(0, \sigma^{2}\bf{I})$.
Further, the symbol vector, $\bm{y}$, is mapped into the arranged feature tensor, $\tilde{\bm{A}}$, and the feature order is restored by the inverse feature arrangement module, $g_2(\cdot)$, i.e.,
\begin{equation}
	\hat{\bm{A}}=g_2(\tilde{\bm{A}}).
\end{equation}

Finally, the receiver decodes the restored feature tensor, $\tilde{\bm{A}}$, to reconstruct the source data, given by
\begin{equation}
	\hat{\bm{s}}=g_1(\tilde{\bm{A}};\bm{\phi}),
\end{equation}
where $\bm{\phi}$ denotes the parameter set of the decoder, $g_1(\cdot)$.
The system loss is defined as
\begin{equation} \label{Loss}
	\mathcal{L}\triangleq d(\hat{\bm{s}}, \bm{s}) =\frac{1}{l}||\hat{\bm{s}}-\bm{s}||^2.
\end{equation}

\subsection{Channel Prediction Module}
To achieve feature arrangement, the CSI in the future period is required.
The channel prediction module keeps sampling the CSI and makes predictions accordingly. The predicted CSI sequence is given by
\begin{equation}
	\tilde{\bm{h}}=\eta_2(\bm{h} ; \bm{\psi_3}),
\end{equation}
where $\bm{h}\in \mathbb{C}^{t_1}$ denotes the sampled CSI sequence, $\tilde{\bm{h}}\in \mathbb{C}^{t_2}$ denotes the predicted CSI sequence, $t_1$ and $t_2$ represent the length of $\bm{h}$ and $\tilde{\bm{h}}$, respectively, and $\bm{\psi_3}$ represents the parameter set of the channel prediction module, $\eta_2(\cdot)$.

Specifically, we consider a time division duplex (TDD) system, where the improved sum-of-sinusoids (SOS) model of \cite{JakeModel} is employed to simulate correlated wide-sense stationary (WSS) Rayleigh fading channels.
In particular, the $n$-th sample of the CSI, $h_n$, is given by
\begin{equation*} \label{SOS}
	h_n = h(n T_s)=\frac{1}{\sqrt{M}} \sum_{m=1}^M[x_{\mathrm{I}, m}(n T_s)+\mathrm{j} x_{\mathrm{Q}, m}(n T_s)],
\end{equation*}
where $T_s$ is the sampling period, $M$ is the number of multipaths, $x_{\mathrm{I}, m}(n T_s)$ and $x_{\mathrm{Q}, m}(n T_s)$ are the $m$-th in-phase component and the $m$-th quadrature component, respectively, which are given as
\begin{subequations}  \label{IQComponents}
	\begin{align*}
		x_{\mathrm{I}, m}(n T_s) & = A_m \cos [(2 \pi f_{\mathrm{D}}^{\max} n T_s+\psi_m) \cos (\alpha_m)+\phi_m], \\
		x_{\mathrm{Q}, m}(n T_s) & = B_m \sin [(2 \pi f_{\mathrm{D}}^{\max} n T_s+\psi_m) \cos (\alpha_m)+\phi_m],
	\end{align*}
\end{subequations}
where $A_m$ and $B_m$ are random attenuations with the distribution, $\mathcal{N}(0, 1)$, $\alpha_m$ and $\phi_m$ denote the arrival angle and the phase shift of the $m$-th path, respectively, and $f_{\mathrm{D}}^{\max}$ is the maximum Doppler shift in Hz.
According to \cite{JakeModel}, the additional phase term, $\psi_m$, is demanded to ensure that the model generates WSS random variables.

\begin{figure}[t]
	\centering
	\includegraphics[width=0.20\textwidth]{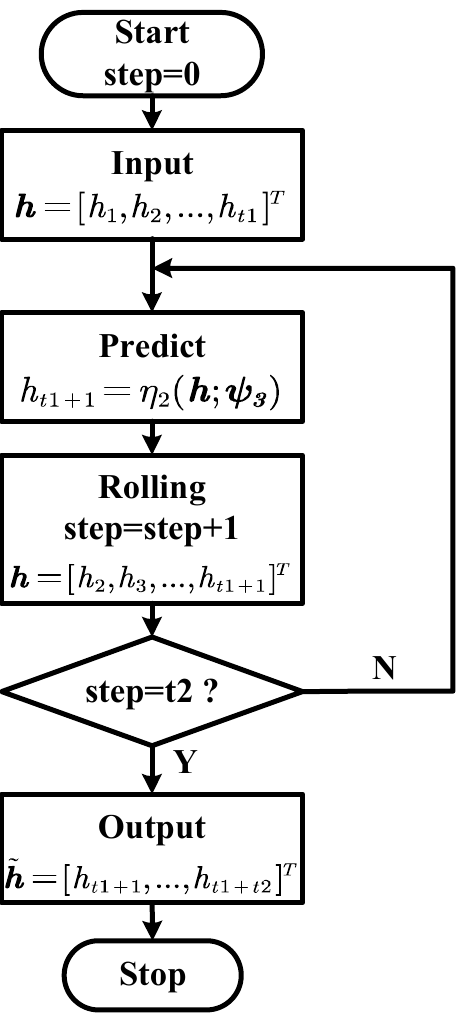}
	\caption{The one-step-ahead rolling forecast method for channel prediction.}
	\label{ChPred}
\end{figure}

Long short-term memory (LSTM) models have excellent performances in time series prediction tasks \cite{LSTM}.
Inspired by this, we design the channel prediction module based on the LSTM network.
Particularly, we adopt the one-step-ahead rolling forecast method for channel prediction, as shown in Fig. \ref{ChPred}.
The detailed procedures are presented as follows:
\begin{itemize}
	\item[(i)] Forecast one time-step of the future CSI, $h_{t1+1}$, according to the sampled CSI sequence, $\bm{h}=[h_1, h_2, ..., h_{t1}]^T$.
	\item[(ii)] Roll one step ahead and update $\bm{h}$ with the predicted CSI, i.e., $\bm{h}=[h_2, h_3, ..., h_{t1+1}]^T$.
	\item[(iii)] Forecast the next CSI value, $h_{t1+2}$, according to the updated $\bm{h}$.
	\item[(iv)] Perform (ii) and (iii) iteratively until  the $h_{t1+t2}$ is forecasted, and the sequence, $[h_{t1+1}, h_{t1+2}, ..., h_{t1+t2}]$, is the predicted sequence.
\end{itemize}

\section{Feature Arrangement}  \label{Methods}
In this section, we elaborate further on the details of the feature arrangement.

\subsection{Feature Priority Module} 
The encoded feature tensor $\bm{A}$ is firstly fed to the feature priority module.
This module determines the priorities of different features, which come out as an output vector, $\bm{\xi}\in \mathbb{R}^c$, where $c$ is the number of features in $\bm{A}$.
The process is expressed as
\begin{equation}
	\bm{\xi}=\eta_1(\bm{A} ; \bm{\psi_1}, \bm{\psi_2}),
\end{equation}
where $\bm{\psi_1}$ and $\bm{\psi_2}$ denote the parameter sets of the networks in this module.

\subsubsection{Feature Priority}
The semantic features have different importance and robustness, where importance indicates how much contribution the feature can provide for the system performance, and robustness indicates its capability to tolerate semantic noise.

Considering transmission in physical channels, the features with high importance or low robustness have higher priority to be transmitted when the CSI is good.
Inspired by this, we propose a metric to quantify the transmission priorities of different features, termed feature priority, which is defined as
\begin{equation}
		\xi = \alpha \cdot w + \beta \cdot(1 - r),
\end{equation}
where $\xi$, $w$, and $r$ denote the feature priority, the feature importance, and the feature robustness, respectively, $\alpha$ and $\beta$ represent the preference for importance and robustness, respectively.
To make the formula meaningful, $w$ and $r$ are both normalized to the same interval, $[0, 1]$.
Moreover, the coefficients, $\alpha$ and $\beta$, are subject to
\begin{subequations}
	\begin{eqnarray}
		\alpha + \beta = 1, \\
		\alpha > 0, \beta > 0.
	\end{eqnarray}
\end{subequations}

\subsubsection{Feature Importance}
We compute the feature importance based on the gradients of the system loss, $\mathcal{L}$, with respect to the features.
The gradients reflect the correlation between the loss and a certain feature, which further indicates how much contribution the feature provides to the system performance. 

Considering the $k$-th feature, $\bm{A}_k\in \mathbb{R}^{h\times w}$, of the feature tensor, $\bm{A}$, we firstly compute the gradients of $\mathcal{L}$ with respect to $\bm{A}_k$, and obtain a gradient matrix, denoted as $\nabla_{\bm{A}_k}\mathcal{L}$, which is then operated by global average pooling.
The obtained value is defined as the importance of the $k$-th feature, given as
\begin{equation}
	w_k=\frac{1}{hw} \sum_{i=1}^{h} \sum_{j=1}^{w} \frac{\partial \mathcal{L}}{\partial a_{k,ij}},
\end{equation}
where $a_{k,ij}$ denotes the element at the $i$-th row and the $j$-th column of the $k$-th feature, $\bm{A}_k$ \cite{GradFI}.
Then, the importance vector of the feature tensor, $\bm{A}$, can be represented as
\begin{equation}
	\bm{w}=[w_1, w_2, ..., w_c]^T.
\end{equation}
The detailed procedures are summarized in Algorithm \ref{FI}.

\begin{algorithm}[t]
	\begin{small}
		\caption{Computing feature importance} 
		\label{FI}
		\DontPrintSemicolon
%		\SetKwInOut{Input}{Input}
%		\SetKwInOut{Output}{Output}
%		\SetKwInOut{Initialize}{Initialize}
		\KwIn{The feature tensor, $\bm{A}\in \mathbb{R}^{c\times h\times w}$, the source data, $\bm{s}$, and the decoder, $g_1(\cdot;\bm{\phi})$.}
		\KwOut{The feature importance vector, $\bm{w}\in \mathbb{R}^c$.}
		Compute the system loss,
		$\mathcal{L} = d(g_1(\bm{A};\bm{\phi}), \bm{s})$.\\
		\For{$k\leftarrow 1$ \KwTo $c$}{
			Compute the gradients of $\mathcal{L}$ with respect to $\bm{A}_k$,
			$\nabla_{\bm{A}_k}\mathcal{L} = \frac{\partial{\mathcal{L}}}{\partial{\bm{A}_k}} = [\frac{\partial{\mathcal{L}}}{\partial{a_{k,ij}}}]$.\\
			Apply average pooling to the gradient matrix,
			$w_k = \frac{1}{hw} \sum\limits_{i=1}\limits^{h} \sum\limits_{j=1}\limits^{w} \frac{\partial{\mathcal{L}}}{\partial{a_{k,ij}}}$.
		}
	\end{small}
\end{algorithm}

\begin{algorithm}[t]
	\begin{small}
		\caption{Computing feature robustness} 
		\label{FR}
		\DontPrintSemicolon
%		\SetKwInOut{Input}{Input}
%		\SetKwInOut{Output}{Output}
%		\SetKwInOut{Initialize}{Initialize}
		\KwIn{The feature tensor, $\bm{A}\in \mathbb{R}^{c\times h\times w}$, the source data, $\bm{s}$, and the decoder, $g_1(\cdot;\bm{\phi})$.}
		\KwOut{The feature robustness vector, $\bm{r}\in \mathbb{R}^c$.}
		Compute the system loss,
		$\mathcal{L} = d(g_1(\bm{A};\bm{\phi}), \bm{s})$.\\
		\For{$k\leftarrow 1$ \KwTo $c$}{
			Generate the semantic noise, $\Delta \bm{\delta}_k^*$, based on (\ref{NoiseProblem}).\\
			Perturb $\bm{A}$ at the $k$-th feature with the zero-padded semantic noise, i.e., $\bm{A} + P(\Delta \bm{\delta}_k^*)$.\\
			Compute the system loss with the perturbed feature tensor,
			$\mathcal{L}' = d(g_1(\bm{A} + P(\Delta \bm{\delta}_k^*);\bm{\phi}), \bm{s})$.\\
			Compute the loss increment,
			$\Delta \mathcal{L} = \mathcal{L}' - \mathcal{L}$.\\
			Compute the reciprocal of $\Delta \mathcal{L}$,
			$r_k = \frac{1}{\Delta \mathcal{L}}$.
		}
	\end{small}
\end{algorithm}

\subsubsection{Feature Robustness}
To compute the feature robustness, we firstly generate semantic noises for each feature, which aims to maximize the system loss, $\mathcal{L}$.
The loss increment caused by adding the generated noise to a certain feature reflects its tolerance to the semantic noise, which indicates its robustness.

Considering the $k$-th feature $\bm{A}_k$, the generation of semantic noise can be modeled as solving the following optimization problem \cite{RobustSem}:
\begin{subequations}  \label{NoiseProblem}
	\begin{eqnarray}
		& \max\limits_{\bm{\delta}_k} & d(g_1(\bm{A} + P(\bm{\delta}_k); \bm{\phi}), \bm{s}) \label{Optim}\\
		& \textrm{s.t.} & \|\bm{\delta}_k \|_2 \leq \epsilon, \label{PowerCons}
	\end{eqnarray}
\end{subequations}
where $\bm{\delta}_k\in \mathbb{R}^{h\times w}$ denotes the semantic noise generated for $\bm{A}_k$, and $P(\cdot)$ is a zero-padding function that pads $\bm{\delta}_k$ into a tensor with the shape of $c\times h\times w$.
The zero-padding operation ensures that the feature tensor $\bm{A}$ is only perturbed at the $k$-th feature and the rest of the features remains the same.
Constraint (\ref{PowerCons}) limits the power of the semantic noise.
To solve this problem, we employ the trust region policy optimization algorithm \cite{TRPO}. 
The algorithm searches the optimal noise, $\bm{\delta}_k^*$, and limits each searching step within a trust region, which ensures that the current step is the optimal before it reaches a local or global optimal solution.
Further, we add the generated semantic noise, $P(\Delta \bm{\delta}_k^*)$, to the feature tensor, $\bm{A}$, and define the robustness of $\bm{A}_k$ as the reciprocal of the loss increment, $\Delta \mathcal{L}$:
\begin{equation}
		r_k\! \triangleq\! \frac{1}{\Delta \mathcal{L}}
		\! = \!\frac{1}{d(g_1(\bm{A} + P(\Delta \bm{\delta}_k^*); \bm{\phi}), \bm{s}) 
	- d(g_1(\bm{A}; \bm{\phi}), \bm{s})},
\end{equation}
where $r_k$ denotes the robustness of $\bm{A}_k$.
Then, the robustness vector of the feature tensor $\bm{A}$ can be represented as
\begin{equation}
	\bm{r} = [r_1, r_2, ..., r_c]^T.
\end{equation}
The detailed procedures are summarized in Algorithm \ref{FR}.

\begin{figure}[t]
	\includegraphics[width=0.45\textwidth]{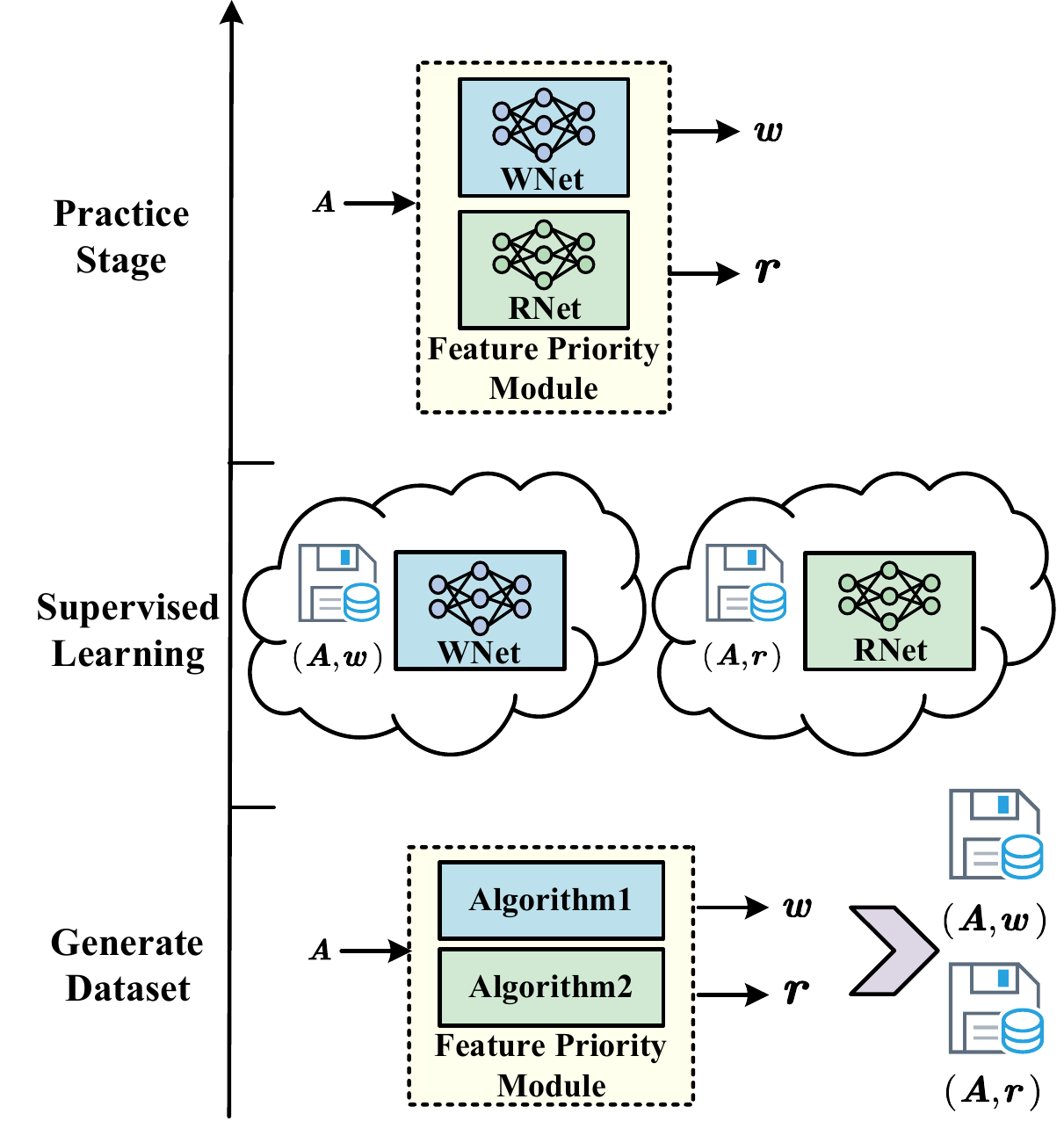}
	\caption{Three stages of knowledge distillation.}
	\label{KD}
\end{figure}

\subsubsection{Knowledge Distillation}
The aforementioned algorithms are still impractical to be performed at the practice stage.
In particular, the system loss, $\mathcal{L}$, is required to be computed in the algorithms.
However, it can only be computed after transmission, while the feature priority is expected to be computed before transmission.
Therefore, we employ the knowledge distillation technique to empower the feature priority module.

The knowledge distillation technique are widely used to transfer the knowledge of a heavyweight network (teacher model) into a lightweight one (student model) \cite{KD}.
Inspired by this, we transfer the knowledge of the algorithms into student models, which can be employed at the practice stage.
Moreover, the student models can also reduce the latency and computational overheads of the feature priority module.

Specifically, we trained two lightweight networks, named WNet and RNet, to simulate the algorithms for computing the feature importance and feature robustness, respectively.
As presented in Fig. \ref{KD}, the distillation process can be summarized into three stages.
The detailed procedures are provided as follows:
\begin{itemize}
	\item[(i)] Compute the feature importance and feature robustness using the aforementioned algorithms. Then create a new dataset to store the yielded importance vector, $\bm{w}$, and the corresponding feature tensor, $\bm{A}$, pair by pair.
	The same goes for the robustness vectors, $\bm{r}$.
	\item[(ii)] Train the WNet and RNet on the generated datasets, respectively.
	\item[(iii)] Substitute the student models for the algorithms in the feature priority module at the practice stage.
\end{itemize}

\subsection{Feature Arrangement Module}
The feature tensor, $\bm{A}$, the feature priority vector, $\bm{\xi}$, and the predicted CSI sequence, $\tilde{\bm{h}}$, are all treated as the input of the
feature arrangement module, $f_2(\cdot)$.
According to the priorities of different features and the future CSI, this module arranges the order of the features in $\bm{A}$.
The arranged feature tensor is given as
\begin{equation}
	\tilde{\bm{A}} = f_2(\bm{A}, \bm{\xi}, \tilde{\bm{h}}).
\end{equation}

\begin{figure}[t]
	\includegraphics[width=0.49\textwidth]{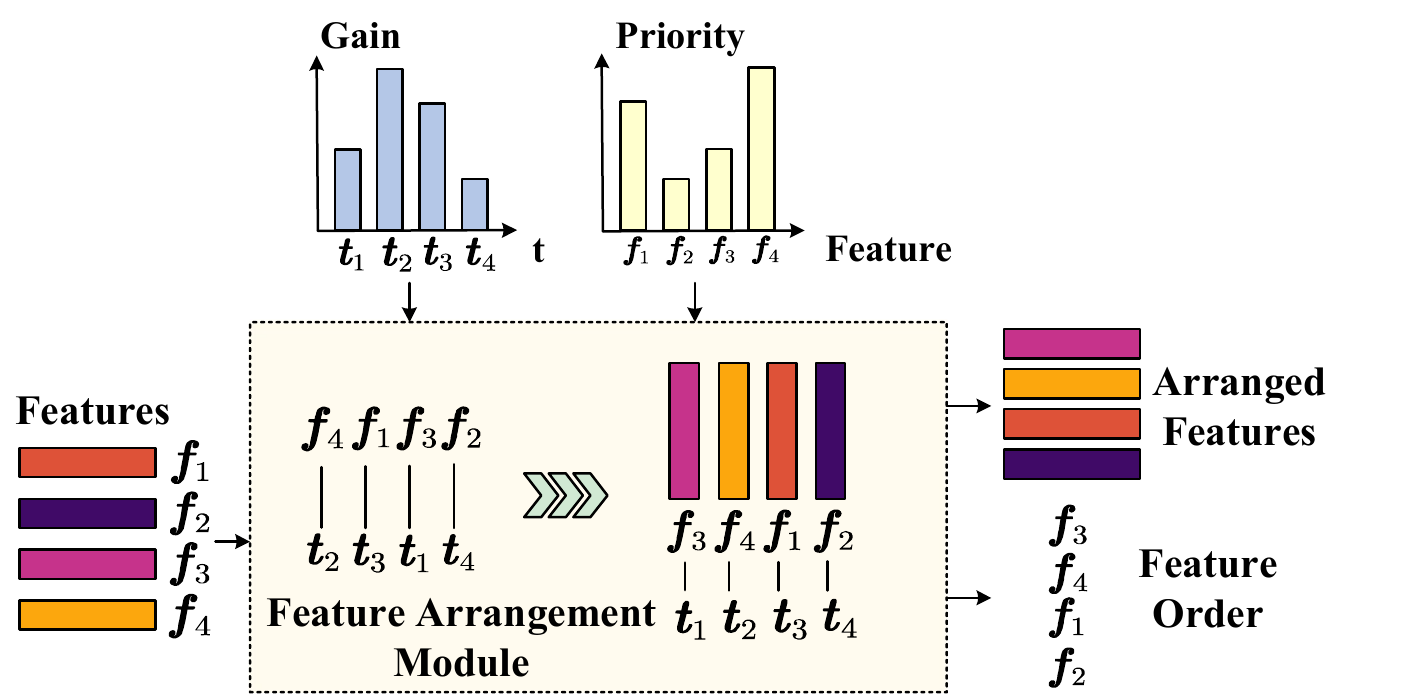}
	\caption{The feature arrangement module.}
	\label{FeaArrMod}
\end{figure}

\begin{algorithm}[t]
	\begin{small}
		\caption{Feature arrangement algorithm} 
		\label{FArrmodule}
		\DontPrintSemicolon
%		\SetKwInOut{Input}{Input}
%		\SetKwInOut{Output}{Output}
%		\SetKwInOut{Initialize}{Initialize}
		\KwIn{The feature tensor, $\bm{A}\in \mathbb{R}^{c\times h\times w}$, the feature priority vector, $\bm{\xi}\in \mathbb{R}^c$, and the predicted CSI sequence, $\tilde{\bm{h}} \in \mathbb{C}^c$.}
		\KwOut{The arranged feature tensor, $\tilde{\bm{A}}\in \mathbb{R}^{c\times h\times w}$, and the feature order, $\bm{\eta}\in \mathbb{N}^c$.}
		Compute the amplitude of each element in $\tilde{\bm{h}}$, i.e., $\tilde{\bm{h}}=abs(\tilde{\bm{h}})\in \mathbb{R}^c$.\\
		Sort $\tilde{\bm{h}}$ and mark each element with its original index, then obtain an index vector, $\bm{u}\in \mathbb{N}^c$.\\
		Sort $\bm{\xi}$ and mark each element with its original index, then obtain an index vector, $\bm{v}\in \mathbb{N}^c$.\\
		\For{$i\leftarrow 1$ \KwTo $c$}{
			$\,\bm{\eta}[\bm{u}[i]] = \bm{v}[i]$.\\
			$\tilde{\bm{A}}[\bm{u}[i]] = \bm{A}[\bm{v}[i]]$.
		}
	\end{small}
\end{algorithm}

In particular, the module firstly operates the feature priority vector, $\bm{\xi}$, and the predicted CSI sequence, $\tilde{\bm{h}}$, through descent sorting, and marks each element with its original index.
Then, the module takes both index vectors and matches their elements pair by pair successively, as shown in Fig. \ref{FeaArrMod}.
This procedure is to arrange the order of the features and assign each feature to the most suitable time slot of transmission.
Subsequently, the arranged feature order is applied to the original feature tensor, $\bm{A}$.
Finally, the module outputs the arranged feature tensor and the feature order.
The feature order is exploited by the inverse feature arrangement module at the receiver side to restore the original feature tensor.
The details of the feature arrangement algorithm are summarized in Algorithm \ref{FArrmodule}.

This design improves the reliability of image transmission and brings remarkable performance gain compared to existing semantic communication systems without feature arrangement.
Furthermore, it also enhances the interpretability of the proposed system.

\section{Simulation Results} \label{Simulation}

\begin{table}\footnotesize
	\centering
	\caption{Settings of the employed networks.}  
	\label{Setting}
	\begin{tabular}{|c|c|c|}
		\hline
		& Layer Name & Dimension  \\
		\hline
		\multirow{3}{*}{\thead{Transmitter \\ (Encoder)}}
		& ConvLayer & 16 (kernels) \\
		\cline{2-3}
		& $3\times$ ConvLayer & 32 (kernels) \\
		\cline{2-3}
		& ConvLayer & 24 (kernels)  \\
		
		\hline
		\multirow{3}*{\thead{Receiver \\ (Decoder)}}
		& $3\times$ TransConvLayer & 32 (kernels) \\
		\cline{2-3}
		& TransConvLayer & 16 (kernels) \\
		\cline{2-3}
		& TransConvLayer & 3 (kernels) \\
		
		\hline
		\multirow{2}*{\thead{Channel Prediction}}
		& 2$\times$LSTM Layer & 50  \\
		\cline{2-3}
		& Dense & 2 \\
		
		\hline
		\multirow{2}*{\thead{WNet}}
		& AvgPooling& 25  \\
		\cline{2-3}
		& $2\times$ Dense & 24  \\
		
		\hline
		\multirow{2}*{\thead{RNet}} 
		& AvgPooling& 25  \\
		\cline{2-3}
		& $2\times$ Dense & 24  \\
		\hline
	\end{tabular}
\end{table}

In this section, we compare the proposed FAST with a basic semantic communication system proposed in \cite{DJSCC}, referred to as DJSCC, under the Rayleigh channel.
We adopt CIFAR-10 as the dataset, which consists of $60,000$ images with the size of $32\times 32\times3$.
By following \cite{DJSCC}, we define the image size, $l$, the channel input size, $k$, and $R = k/l$ as the source bandwidth, the channel bandwidth, and the bandwidth ratio, respectively.
The encoder and decoder are trained at the bandwidth ratio $R = 1/4$ and $\mathrm{SNR_{train}}=$ $7$ dB, $13$ dB, $19$ dB as $3$ different system models.
Note that the structure of the encoder and decoder is the same between FAST and DJSCC.
Moreover, they are both tested at $\mathrm{SNR_{test}}$ from $0$ dB to $25$ dB.
The WNet and RNet in FAST are trained independently.
The settings of the employed networks are presented in Table \ref{Setting}.

The performance of the proposed FAST and the benchmark, DJSCC, is quantified in terms of peak signal-to-noise ratio (PSNR), which is defined as
\begin{equation}
	\mathrm{PSNR}=10 \log _{10} \frac{\mathrm{MAX}^2}{\mathrm{MSE}}(\mathrm{dB}),
\end{equation}
where $\mathrm{MSE}=\frac{1}{l}\|\bm{s}-\hat{\bm{s}}\|^2$ and $\mathrm{MAX}$ is the maximum possible value of the image pixel.
We offer the performance of the following schemes:
\begin{itemize}
	\item PC+FP+KD: The proposed FAST scheme. The CSI is obtained via channel prediction (PC), and the feature priority (FP) module is improved based on the knowledge distillation (KD) technique.
	\item KC+FP+KD: An ideal variant of FAST assuming precisely known future CSI (KC).
	\item KC+FP: A variant of FAST assuming precise CSI without employing the knowledge distillation technique.
	\item PC+FP: A variant of FAST without the knowledge distillation technique.
	\item DJSCC: A basic semantic communication system without feature arrangement.
\end{itemize}

\begin{figure}[t]
	\begin{centering}
		\includegraphics[width=0.4\textwidth]{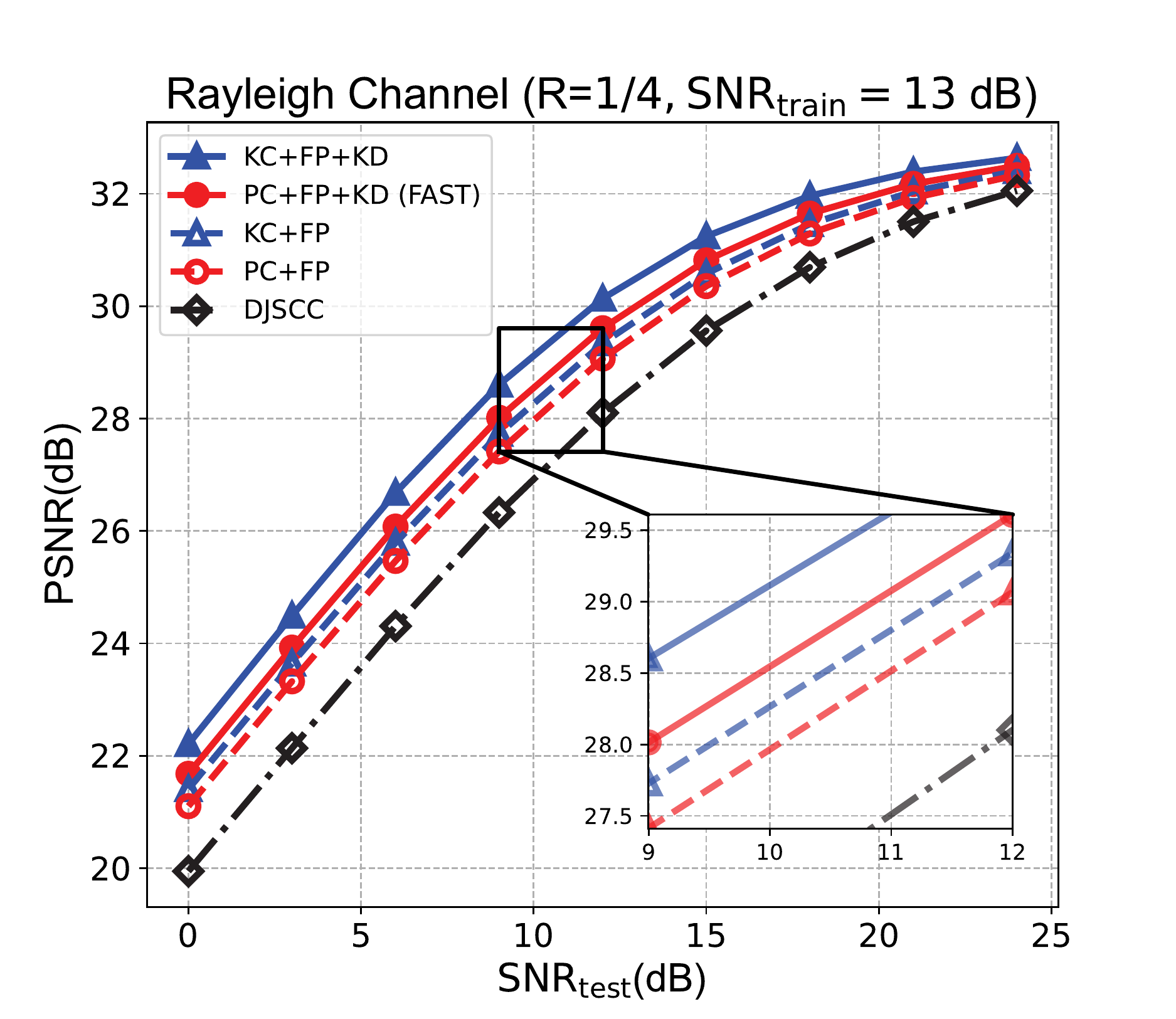}
		\par\end{centering}
	\caption{The performance of different schemes versus SNR.}
	\label{Comparison}
\end{figure}

Fig. \ref{Comparison} shows the performance of different schemes versus SNR.
It is readily seen that the FAST and all its variants significantly outperform the benchmark, especially at low SNR regimes.
It is mainly because the FAST manages to transmit the features with high priority when the CSI is good.
Besides, the schemes with predicted CSI perform worse than the schemes that assume the CSI is precisely known.
It is because the channel prediction module cannot ensure precise prediction and the resulting performance loss is inevitable.
However, the performance of the schemes with precise CSI can be approached by improving the accuracy of channel prediction.
Furthermore, the schemes with the knowledge distillation technique outperform those without the knowledge distillation technique.
It is because the generalization of networks makes it possible for the student models to perform even better than the algorithms.

\begin{figure}[t]
	\begin{centering}
		\includegraphics[width=0.4\textwidth]{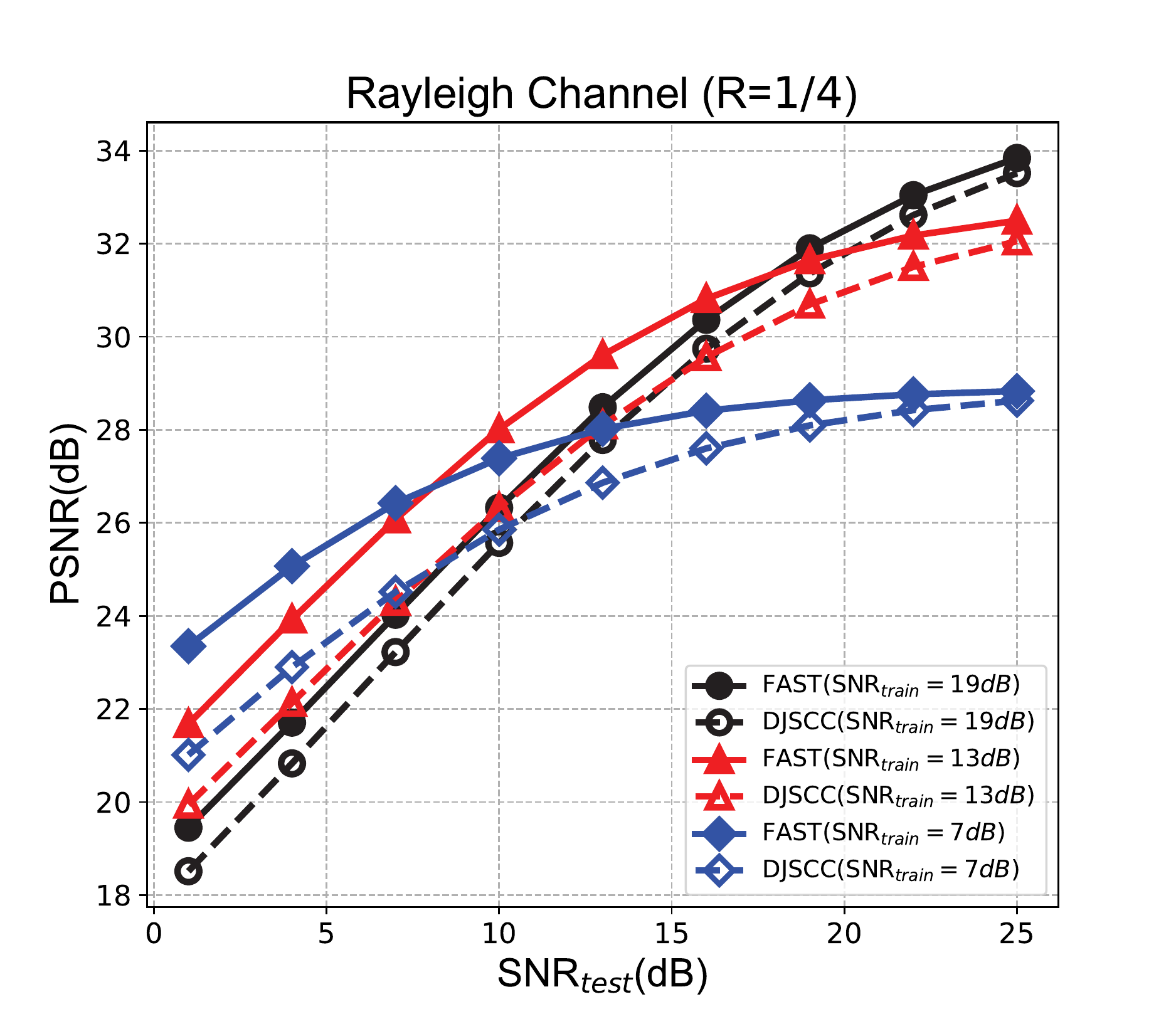}
		\par\end{centering}
	\caption{Comparison between the FAST and DJSCC at different training SNRs.}
	\label{FinalPerform}
\end{figure}

Fig. \ref{FinalPerform} illustrates the performance of the FAST and DJSCC at different training SNRs.
We can observe that all three FAST models trained at different SNRs outperform the corresponding DJSCC model over the entire $\mathrm{SNR_{test}}$ region, which demonstrates that the proposed FAST can maintain its superiority at different training SNRs.
Moreover, the performance gain of the FAST trained at low SNR is larger than that trained at high SNR.
It is because the feature arrangement scheme significantly mitigates the performance degradation caused by the corruption of high-priority features, especially at low SNR regimes.
This result exhibits the advantages of the proposed FAST under harsh channel conditions.

\section{Conclusion} \label{Conclusion}
In this paper, we have proposed a novel semantic communication system with feature arrangement to improve the performance of image transmission.
Particularly, we aim to transmit the prior features under better CSI.
To this end, a novel algorithm has been proposed to calculate the priority of different features.
Further, the feature arrangement module has been developed to schedule the transmission order of different features, based on the feature priority and the predicted CSI. 
Simulation results have shown that the proposed scheme significantly improves the performance of image transmission.

\bibliographystyle{IEEEtran}
\bibliography{IEEEabrv,Semantic}

\end{document}